\documentstyle[11pt,cs11,html,epsf]{article}

\begin{document}

\title{Multiwavelength optical observations of chromospherically 
active binary system OU Gem}

\author{A. Latorre, 
D. Montes, 
M.J. Fern\'andez-Figueroa}
\affil{Departamento de Astrof\'{\i}sica, 
U. Complutense de Madrid, Spain}

% Notice that some of these authors have alternate affiliations, which
% are identified by the \altaffilmark after each name.  The actual alternate
% affiliation information is typeset in footnotes at the bottom of the
% first page, and the text itself is specified in \altaffiltext commands.
% There is a separate \altaffiltext for each alternate affiliation
% indicated above.

% The nice thing about this method is that it saves space on the first page!

% BUT if you've used \altaffiltext and you think you'll be adding 
% *footnotes*, then you need to update the footnote counter!

\setcounter{footnote}{3}

% The abstract is entered in a LaTeX "environment", designated with paired
% \begin{abstract} -- \end{abstract} commands.  Other environments are
% identified by the name in the curly braces.

\begin{abstract}

 We have analysed several multiwavelength optical observations 
(H$\alpha$, H$\beta$, H$\epsilon$, Ca~{\sc ii} H \& K, Ca~{\sc ii} IRT
lines) of chromospherically active binary system OU Gem, taken in
 three observation runs: March 1996, April 1998 (using the SOFIN 
spectrograph and the Nordic Optical Telescope) and January 1999 
(using the ESA-MUSICOS spectrograph and the Isaac Newton Telescope).
The spectra show double H$\alpha$ and H$\beta$ absorption lines, 
but after applying the spectral subtraction technique, clear excess 
emission is seen in both components. The two stars have the Ca~{\sc ii} 
IRT lines
in emission, strong emission in the Ca~{\sc ii} H \& K lines and 
small emission in the H$\epsilon$ line. Although they have very 
similar spectral types, 
the cool one is more active than the hot one in all runs, 
except for MUSICOS run, and in all spectral lines,
except for Ca~{\sc ii} IRT lines.  
%The biggest difference in the excess emission equivalent width between
%both components is found in January 1999.
\end{abstract}

% Keywords should be included, but they are not printed in the hardcopy.
% They will be used by the Editors to help organize poster papers by
% category though!


% That's it for the front matter.  On to the main body of the paper.
% We'll only put in tutorial remarks at the beginning of each section
% so you can see entire sections together.

% 
% OK - to make things easier for the Editors, we're going to put
% all of our object aliases up front since we only have to declare
% them once in the paper.  Some people prefer to use NGC 7078 for
% M 15, but we like good old Messier, so that's what we'll index by.
% But we'll cross-reference it here so that people who do like NGC 7078
% won't have to remember that it's also M 15!
%
% Remember - we identify objects by putting an asterisk in front of the name!
%
\index{*OU Gem|HD 45088}

\section{Introduction}

  OU Gem (HD 45088) is a bright (V= 6.79, Strassmeier et al.~1990) 
and nearby (d= 14.7 pc, ESA 1997) BY Dra-type
SB2 system (K3V/ K5V) with a 6.99-day orbital period and a noticeable
eccentricity (Griffin $\&$ Emerson~ 1975). Both components show 
Ca~{\sc ii}~H~$\&$~K
emission, though the primary emission is slightly stronger
than the secondary emission. The H$\alpha$ line is in absorption for the
primary and filled-in for the secondary (Bopp~1980, Bopp et al. 1981a, b,
 Strassmeier et al. 1990). Dempsey~et~al.~(1993) observed that the 
Ca~{\sc ii} IRT lines were filled-in.
It is interesting that the orbital and rotational
periods differ in 5\% due to the appreciable orbital
eccentricity (e= 0.15), according to Bopp (1980).

 Although BY Dra systems are main-sequence stars, their evolutionary 
stage is not clear. OU Gem has been listed by Soderblom et al. (1990) 
and Montes et al. (1999) as a possible member of the UMa moving 
group (300~Myr), and this suggests it may be a young star.

In this paper we present multiwavelength optical observations of 
chromospherically active binary system OU Gem.

%----------------------------------
\section{Observations}

 Spectroscopic observations in several optical chromospheric activity
indicators of OU Gem and some inactive stars of similar spectral type and
luminosity class have been obtained during three observation runs.
\newline
1) Two runs were carried out with the 2.56~m Nordic Optical Telescope
(NOT) at the Observatorio del Roque de Los Muchachos (La Palma, Spain) in
March 1996 and April 1998, using the SOFIN echelle spectrograph covering
from 3632~\AA$\ $ to 10800~\AA$\ $
 (resolution from $\Delta$$\lambda$ 0.15 to 0.60 \AA), with the
1152$\times$770 
 pixels EEV P88200 CCD detector.
\newline
2) The last run was carried out with the 2.5~m INT at the Observatorio
del Roque de Los Muchachos (La Palma, Spain) in January 1999 using 
Multi-Site Continuous Spectroscopy (MUSICOS), covering from 3950~\AA$\ $ to
9890~\AA$\ $  (resolution from $\Delta$$\lambda$ 0.15 to 0.40 \AA), 
with the 2148$\times$2148 pixels
SITe1 CCD detector.

%----------------------------------
\begin{figure}
\plotone{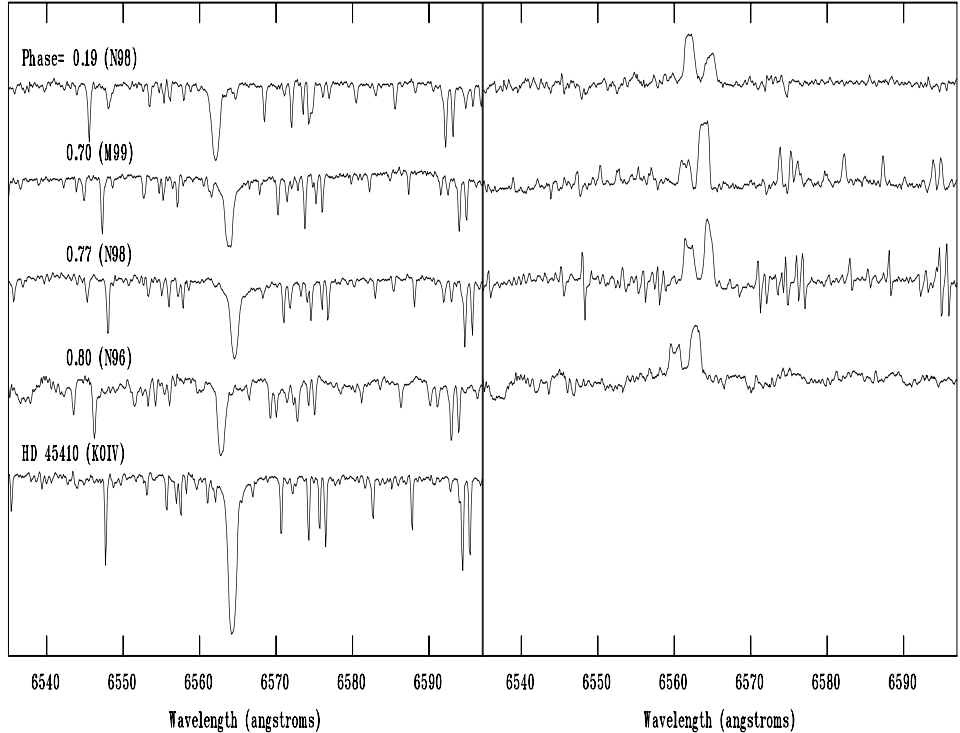}
\caption{The left panel shows the OU Gem observed spectra and the 
HD 45410 (K0IV) 
reference star, in the H$\alpha$ line region.
The right panel shows the OU Gem subtracted spectra.} \label{fig-1}
\end{figure}
%----------------------------------
 
       In the three runs we obtained 4 spectra of OU Gem at different
orbital phases.
%In Table 1 we give the observing log. For each
%observation we list date, UT, orbital phase ($\varphi$) and the signal to noise
%ratio (S/N). Table 2 shows the 
The stellar parameters of OU Gem have been adopted from
Strassmeier et al. (1993), except for distance (given by the Hipparcos
catalogue (ESA 1997)), the B-V color index and the radius (taken from 
the spectral type).
        The spectra have been extracted using the standard reduction
procedure of the IRAF package (bias subtraction, flat-field division and
optimal extraction of the spectra). The wavelength calibration was
obtained by taking spectra of a Th-Ar lamp, for NOT runs, and a Cu-Ar lamp, 
for INT runs. Finally the spectra have
been normalized by a polynomial fit to the observed continuum.
        The chromospheric contribution in the activity indicators has been
determined using the spectral subtraction technique (Montes et al. 1995a, b).

% Let's make a gratuitous index entry for Lick Observatory, under
% Observatories, Lick
\index{Observatories!Roque de Los Muchachos}

% In this section, we see the use of the \subsection command to set off
% an independent subsection.  We only have one here; usually there would
% be several.
%
% We show the use of several of the displayed math environments described
% in the User Guide, and you get a healthy dose of mathematical typesetting
% examples.  Also, observe the use of the LaTeX \label command after the
% \subsection to give a symbolic tag to the subsection for cross-referencing
% in a \ref command.  LaTeX automatically numbers the sections, equations,
% tables, etc. as it goes, so in general you don't know what number something
% is going to have.  We'll refer to the "hairymath" section a little later.

%----------------------------------
\begin{figure}
\plotone{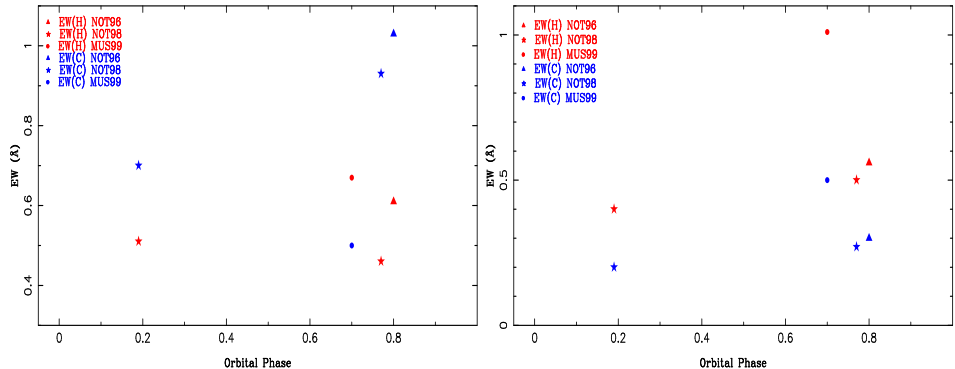}
\caption{The left panel shows excess H$\alpha$ emission equivalent width 
versus the orbital phase for the hot component (H) and for the cool 
component (C) in all observation runs. The right panel shows the same for 
the Ca~{\sc ii}~$\lambda$8662 line.} \label{fig-2}
\end{figure}
%----------------------------------

%----------------------------------
\section{The H$\alpha$ line}

 We have taken several spectra of OU Gem in the H$\alpha$ line region  in
three different runs and at different orbital phases.
The line profiles are displayed in Fig.~\ref{fig-1}. For each observation,
we have plotted the
 observed spectrum (left panel) and the subtracted spectrum (right panel).
In the observed spectra, we can see an absorption line for the primary star and 
a nearly complete filling-in for the secondary star. After applying the 
spectral subtraction technique, clear excess H$\alpha$ emission is obtained 
for the two components, being stronger for the hot one.
The excess H$\alpha$ emission equivalent width (EW) is measured in the
subtracted spectrum and corrected for the contribution of the components
to the total continuum.
%The excess H$\alpha$ emission EW is given in Table
%3 (column 3), 
 We took one spectrum in this region in Dec-92 (Montes et al. 1995b). 
At the orbital phase of this observation ($\varphi$=~0.48) we could not 
separate the emission from both components and we measured the total excess 
H$\alpha$ emission EW. We obtained a similar value to Mar-96, Apr-98 and 
Jan-99 values obtained adding up the excess emission EW from the two 
components. In Fig.~\ref{fig-2} we have plotted the excess H$\alpha$ 
emission EW versus the orbital phase.

%----------------------------------
%\begin{figure}
%
%\plotone{fig5.ps}
%
%\caption{As in Fig. 1 for the H$\beta$ line region.}
% \label{fig-3}
%\end{figure}
%----------------------------------

\section{The H$\beta$ line}

\begin{figure}
\plotone{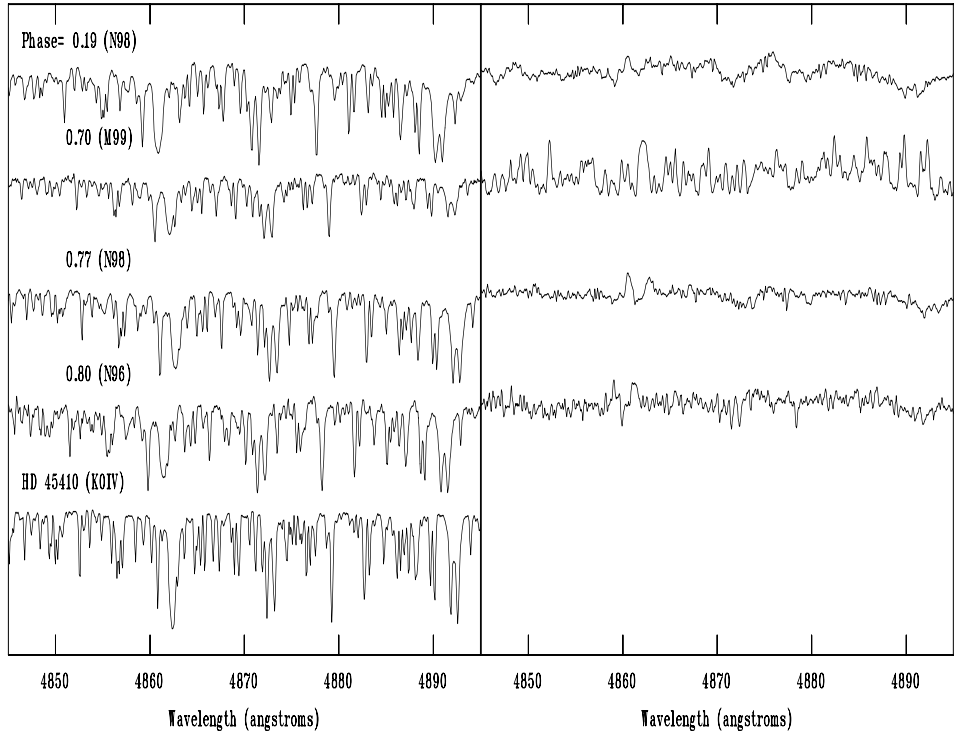}
\caption{As in Fig. 1 for the H$\beta$ line region.}
 \label{fig-3}
\end{figure}

   Four spectra in the H$\beta$ region are available, in three different
runs and at different orbital phases. Looking at the observed spectra in 
Fig.~\ref{fig-3}, we can only see the H$\beta$ line of the primary, 
in absorption. After applying the spectral subtraction tecnique,
small excess H$\beta$ emission is obtained for the two components.
We have determined the excess H$\beta$ emission EW in the subtracted spectra,
the ratio of excess H$\alpha$ and H$\beta$ emission EW,  
and the  $ \frac{ E_{H\alpha} } { E_{H\beta} } $ relation:

\begin{equation}
\frac{ E_{H\alpha} }{ E_{H\beta}}=  \frac{ EW_{sub} (H\alpha) }{ EW_{sub} 
(H\beta) } \cdot 0.2444 \cdot 2.512^{(B-R)}
\end{equation}

given by Hall \& Ramsey (1992) as a diagnostic for discriminating between
the presence of plages and prominences in the stellar surface.  We have
obtained high $ \frac{ E_{H\alpha} } { E_{H\beta} } $ values for the two 
components, so the emission can come from prominences. 

%----------------------------------
%----------------------------------

\section{The Ca~{\sc ii} IRT lines}

 The Ca~{\sc ii} infrared triplet (IRT) $\lambda$8498, $\lambda$8542, 
and $\lambda$8662 lines are
%----------------------------------
\begin{figure}
\plotone{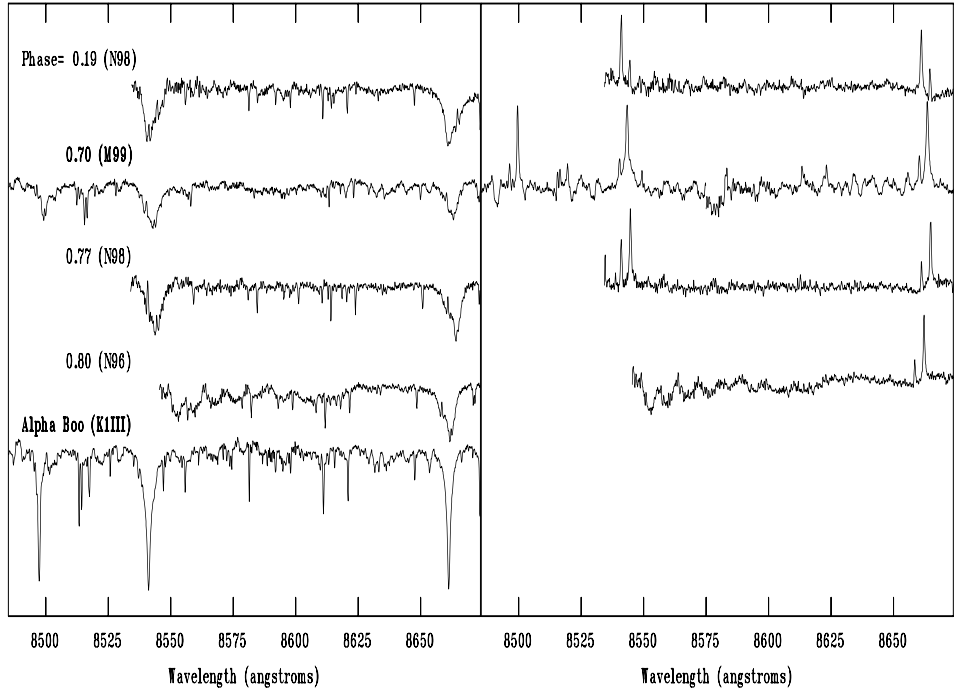}
\caption{As in Fig. 1 for the Ca~{\sc ii} IRT lines region.}
 \label{fig-4}
\end{figure}
%----------------------------------
other important chromospheric activity indicators. We have taken several
spectra of OU Gem in the Ca~{\sc ii} IRT lines region in three different
runs and at different orbital phases, the three Ca~{\sc ii} IRT lines are
only included in MUSICOS 99 observation run. In the observed spectra, 
we can see
both components of OU Gem show these lines in emission superimposed to the 
corresponding absorption (Fig.~\ref{fig-4}).
After applying the spectral subtraction tecnique, clear excess emission 
appears for the two components, being stronger for the hot one. 
In Fig.~\ref{fig-2}  we have plotted
the Ca~{\sc ii} $\lambda$8662 
EW versus the orbital phase for the hot and cool components.

%----------------------------------
%\begin{figure}
%
%\plotone{latorrea5.ps}
%
%\caption{As Fig. 1 for the Ca~{\sc ii} IRT lines region.}
% \label{fig-4}
%\end{figure}
%----------------------------------

\section{The Ca~{\sc ii} H $\&$ K and H$\epsilon$ lines}

\begin{figure}
\plotone{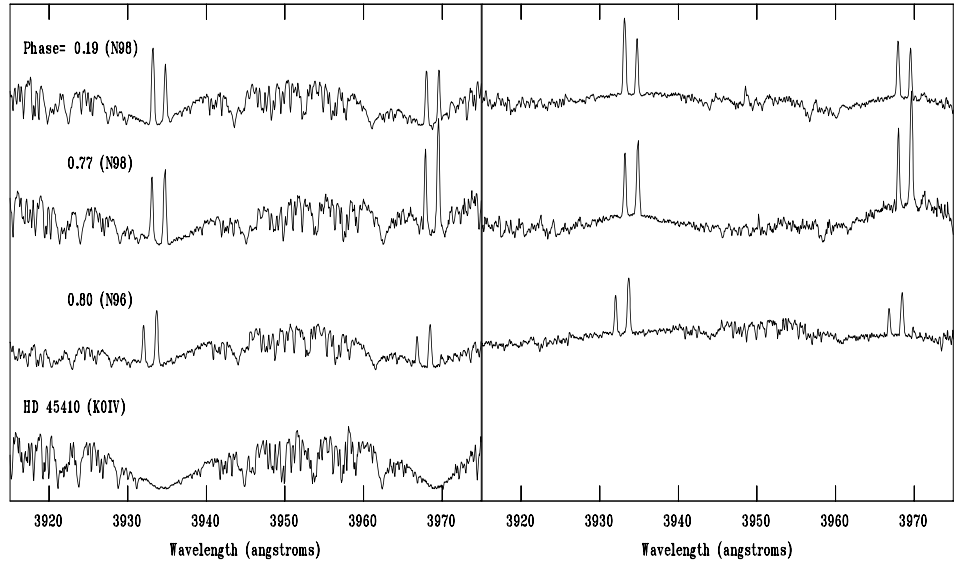}
\caption{As in Fig. 1 for the Ca~{\sc ii} H \& K line region.}
 \label{fig-5}
\end{figure}

We have taken three spectra in the Ca~{\sc ii} H \& K lines region during 
the two NOT (96 \& 98) observation runs.
In Fig.~\ref{fig-5} we can observe that both components of this binary have 
the Ca~{\sc ii} H $\&$~K and H$\epsilon$ lines in emission. We can also see 
that the hot star's excess Ca~{\sc ii} H \& K emission
is bigger than the cool star's. At 0.19 orbital phase, 
we only see the H$\epsilon$ line of the cool component because the H$\epsilon$ 
line of the hot star is overlapped with the Ca~{\sc ii}~H line of the cool one.
 At the other orbital phases, we can only see the H$\epsilon$ line of the hot 
component because the H$\epsilon$ line of the cool one is overlapped with the 
Ca~{\sc ii}~H line of the hot star. 
We took one spectrum in this region in Mar-93 (Montes et al. 1995a, 1996), 
with the 2.2~m telescope at the German Spanish Astronomical
Observatory (CAHA).  At the orbital phase of this observation ($\varphi$= 0.47)
 we could not separate the emission from both components and we measured the 
total excess Ca~{\sc ii}~K~$\&$~H emission EW. We obtained similar values to 
Mar-96 and Apr-98 values obtained adding up the excess emission EW from the 
two components, though we note that the Apr-98 values are slightly bigger than 
the others.

%----------------------------------
%\begin{figure}
%
%\plotone{fig1.ps}
%
%\caption{As Fig. 1 for the Ca~{\sc ii} H \& K line region.}
% \label{fig-5}
%\end{figure}
%----------------------------------

%----------------------------------

% Finally, we have a little acknowledgements section.

\acknowledgments

This work was supported by the Universidad Complutense de Madrid
and the Spanish Direcci\'{o}n General de Ense\~{n}anza Superior e 
Investigaci\'{o}n Cient\'{\i}fica (DGESIC) under grant PB97-0259.

%----------------------------------
%----------------------------------

\end{document}